Title: Quantum mechanical motion of off-center ion in external magnetic field

Author: Mladen Georgiev (Institute of Solid State Physics, Bulgarian Academy of Sciences, 1784 Sofia, Bulgaria)

We consider the magnetostatic response to an external magnetic field of a crystal containing off-center ions, such as $Li^+$ in KCl and KBr or the apical oxygens O(A) in the $La_{2-x}Sr_xCuO_4$ family of layered perovskites. Magnetic dipoles are deduced from the matrix elements of the energy operator of a spinning particle in a magnetic field which particle also satisfies the nonlinear Mathieu equation. The magnetic moments are found to increase in magnitude as the system goes from the lowest energy ground state to the higher lying excited states. Implications for the crystal conductivity by vibronic polarons are also mentioned.

## 1. Foreword

Substitutional ions in crystals, whether host or impurity ones, go off-center as the inversion symmetry breaks up at their crystalline sites (parity violation). It seems commonly accepted nowadays that the off-center (off-site) displacement is produced by the vibronic mixing of a pair of even- and odd- parity electronic states at the ionic site by an odd-parity vibrational mode of the ions at that site [1]. What is needed is a coupling of the electronic states to the vibrational mode and various models have introduced linear or higher-order coupling terms to account for the off-center effect. With the coupling of electronic and vibrational subsystems at hand, the off-center displacement occurs if the coupling is sufficiently strong. Depending on the site-symmetry of the point group, the off-center displacements form a configurational structure which comprises the framework of allowed displacements [2]. Depending on the temperature, the displaced ions perform tunneling or classical rotary transitions between allowed off-center sites which rotations restore on the average the broken inversion symmetry at the normal lattice site. Off-center ions are smeared within the volume of the allowed transition structure, and, given the site-symmetry, the smeared off-center particle may be spherical-like or highly symmetric in shape, but also oblate or prolate if the symmetry and rotational axes are perpendicular or parallel to each other, respectively. Consequently, the resulting shape and possibly migration of the off-center entity may display "barbaronic features" by exhibiting an increased penetrating ability across a crystalline medium despite obstacles.

The "Barbarons" are fancy creatures having both a high cohesion and a high flexibility which help them move across the environment undergoing little, if any, resistance from the obstacles within. The unusual barbaronic features come from both their cohesive power, which makes them stable against tearing apart, and flexibility, which helps them assume any conceivable shape. In reality, both barbaronic features may be displayed by a Van-der-Waals (VdW) structure within a crystalline medium, provided the dispersive cohesive force is sufficiently powerful, such as suggested earlier for a "polaronic matter", due to the polaron gap narrowing and the associated enhanced polarizability [3].

Arising from the vibronic mixing, the off-center ions are the atomic-scale distortions associated with vibronic polarons by Jahn-Teller or Pseudo-Jahn-Teller cooperative interactions [4]. We regard the vibronic polaron as a species migrating across a crystal with enhanced barbaronic permeability. This could have a profound effect on the electric conduction of crystals containing vibronic polarons. It should be noted that the barbaronic features will persist at temperatures sufficiently low that the WdV pairing energy be superior to the thermal energy which will tend to break the bond down.

From a classic viewpoint, the distribution of a displaced ion within the off-center volume is barrier-dependent. There are two essential barriers: One is the formation barrier of off-center to on-center transition, the other is the rotational barrier of a displaced ion atop the off-center structure. Now if the formation barrier is superior to the rotational barrier, the off-center entity will be distributed mainly along the brim of the sombrero potential, while in the opposite case the entity will smear uniformly within the whole volume [5]. In the former case there will be circular currents along the sombrero ring, whereas the brim currents will be largely suppressed in the latter case. The circular motion will give rise to magnetic dipoles, by virtue of the classic Ampere theorem [6]. Undoubtedly the magnetic pair interaction of two ring dipoles will be attractive if they are anti-parallel, or repulsive vice versa. Ultimately, this may give rise to various forms of magnetic behavior, such as anti-ferromagnetism, ferromagnetism, etc. In any event, to study the matter thoroughly, a quantum-mechanical investigation is mandatory to complement the earlier classic étude. We attempted an approach based on Landau's study of the electron spin [7]. This is followed by an extension to the off-center rotation with its specific requirements [8].

## 2. Schrödinger's equation

### 2.1. Free spinning particle

The Schrödinger equation of a spinning particle in an external magnetic field $\mathbf{B} = \mu\, \mathbf{H}$ ($\mu$ is the permeability) for a medium with $\mu = 1$ has been discussed in Landau & Lifshits' book [7]. The relevant Hamiltonian is:

$$\hat{H} = (1/2M)[\boldsymbol{p} - (e / c)\mathbf{A}]^2 - \beta \mathbf{s.H} + U(x,y,z) \tag{1}$$

where $\boldsymbol{p} = -i\hbar\nabla$ is the particle-momentum operator, M is the mass, U(x,y,z) is the electrostatic potential. $\beta\mathbf{s.H}$ is the coupling term to the particle spin. $\mathbf{A} = \mathbf{A}(x,y,z) = ½\, \mathbf{H} \times \mathbf{r}$ is the vector potential of the magnetic field. For an axial field $\mathbf{H}$ parallel to the z-axis, the wave equation turns into

$$\{(1/2M)[\,[\,p_x + (e / 2c)\,H\,y\,]^2 + [\,p_y - (e / 2c)\,H\,x\,]^2 + p_z^2\,] - \beta s_z H + U(x,y,z)\}\Psi = E\,\Psi \tag{2}$$

At U = 0 the operator part to the left does not contain the z coordinate. As a result, $p_z$ commutes with the Hamiltonian and the wave function of a free particle has the form $\psi = \exp((i / \hbar) p_z z)\, \chi(x,y)$. The relevant wave equation for $\chi(x,y)$ obtains as $\psi$ is substituted for in (2):

$$\chi'' + (2M / \hbar^2)\{E + \beta\sigma H - p_z^2 / 2M -$$

$$\tfrac{1}{2} M (e H / M c)^2 (x - x_0)^2 - \tfrac{1}{2} M (e H / M c)^2 (y - y_0)^2\}\chi = 0 \qquad (3)$$

with $x_0 = cp_y /eH$ , $y_0 = - cp_x /eH$, $\sigma = s_z$ . This is the form of the wave equation of two uncoupled though displaced linear harmonic oscillators of total eigen energy $E + \beta\sigma H - p_z^2/2M$ and frequency $\Omega_H = |e|\, H / Mc$, the precession frequency of a gyrating particle. Consequently the eigen energy of a free spinning particle is

$$E = (n_x + \tfrac{1}{2}) ( |e|\, \hbar / M c) H + (n_y + \tfrac{1}{2}) ( |e|\, \hbar / M c) H + p_z^2 / 2M - \beta\sigma H \quad (4)$$

The corresponding eigen function of that particle is

$$\psi = \exp ((i / \hbar) p_z z ) \exp[-(eH / 2e \hbar)(x - x_0)^2]\, H_{nx} [\sqrt{(eH / e \hbar)}(x - x_0)] \times$$

$$\exp[-(eH / 2e \hbar)(y - y_0)^2]\, H_{ny} [\sqrt{(eH / e \hbar)}(y - y_0)] \qquad (5)$$

$H_{nx}$ and $H_{ny}$ are Hermite polynomials. The eigen energy accounted for by the first terms in (4) is one of motion upon the (x,y) plane. Classically, this is the motion along a ring about a fixed center. The quantities $x_0$ and $y_0$ are the analogues of the ring center classical coordinates. Both $x_0$ and $y_0$ are conserving, for each of them commutes with the Hamiltonian. However, the operators $x_0$ and $y_0$ are non-commuting, so that they are not observable simultaneously.

## 2.2. Reorientating off-center ion in crystal

To apply the above line of reasoning to the reorientational motion in-plane of an off-center ion in crystal placed in a magnetic field, we incorporate a finite electrostatic energy U(x,y,z) and set $p_z = z \equiv 0$. At $U \neq 0$ Schrödinger's equation of a rotating off-center ion in crystal reads (Devonshire potential) [2]:

$$\{(1/ 2M)\, \mathbf{p}^2 + (M\omega^2 / b)\, [(d_c - d_b) \sum(x^4 + y^4) + d_b r_0^4] \pm$$

$$E_{JT} [1\pm2 - (E_{\alpha\beta} / E_{JT})^2 ]\}\phi = E\phi \qquad (6)$$

where b is the linear electron-vibrational mode coupling constant, $d_b$ and $d_c$ are the principal-diagonal and side-diagonal third-order coupling constants, $\omega$ is the vibrational frequency renormalized by the electron-mode coupling, $r_0$ is the off-center radius, $E_{\alpha\beta}$ and $E_{JT}$ are the electronic inter-level energy gap and the Jahn-Teller energy involved in the off-center displacements. When the local rotation is in the equatorial plane, the solutions of the above equation are composed of Mathieu's transcendent functions [9]. Some of Mathieu's eigen energies and eigen functions are shown in Figures 1 and 2. We remind that they are relevant to describing the hindered rotation of particles in-plane.

The complete Hamiltonian of a rotating off-center ion in an axial magnetic field will therefore read:

$$\hat{H} = (1/2M)\,[\,[p_x + (e/2c)\,By]^2 + [p_y - (e/2c)\,Bx]^2\,] - \beta\sigma H +$$
$$(M\omega^2/b)\,[(d_c - d_b)\,\Sigma(x^4 + y^4) + d_b\, r_0^4] \pm E_{JT}\,[1 \pm 2 - (E_{\alpha\beta}/E_{JT})^2\,] \quad (7)$$

At H = B = 0 this Hamiltonian is similar though not identical to the Hamiltonian of free rotation along the off-center ring of an off-site ion in crystal. The ion will rotate unimpeded if only linear electron-mode coupling is accountable. As higher cubic terms are incorporated, barriers will appear at reorientational sites pre-given by symmetry along the off-center ring to hinder the free rotation [10].

We solve for Schrödinger's equation based on (7) in terms of a linear combination of harmonic-oscillator eigen functions of a free rotation and Mathieu's eigen functions of a hindered rotation: $\Psi = C_\chi\, \chi + C_\phi\, \phi$, respectively. Clearly, the $(\chi,\phi)$ basis involves states of a merely spinning particle as well as of one undergoing an off-centered orbital motion in addition to spinning. We get a secular equation for the eigen energy E:

$$(<\phi\,|\,\hat{H}\,|\,\phi> - E\,)\,(<\chi\,|\,\hat{H}\,|\,\chi> - E\,) - <\chi\,|\,\hat{H}\,|\,\phi><\phi\,|\,\hat{H}\,|\,\chi> = 0 \quad (8)$$

with the following roots:

$$E = \tfrac{1}{2}\,\{\,(<\phi\,|\,\hat{H}\,|\,\phi> + <\chi\,|\,\hat{H}\,|\,\chi>\,) \pm \surd\,[(<\phi\,|\,\hat{H}\,|\,\phi> - <\chi\,|\,\hat{H}\,|\,\chi>\,)^2 +$$
$$4<\chi\,|\,\hat{H}\,|\,\phi><\phi\,|\,\hat{H}\,|\,\chi>\,]\,\} \quad (9)$$

From the complete equation (9) we get the energy to zeroth-order perturbation, as we discard the off-diagonal matrix elements:

$$E_{0+} \sim\ <\phi\,|\,\hat{H}\,|\,\phi>$$

$$E_{0-} \sim\ <\chi\,|\,\hat{H}\,|\,\chi>, \quad (10)$$

while the energy to first-order perturbation at a small though finite off-diagonal matrix is:

$$E_{I\pm} \sim E_{0\pm}\ \pm$$
$$<\chi\,|\,\hat{H}\,|\,\phi><\phi\,|\,\hat{H}\,|\,\chi>/\,(<\phi\,|\,\hat{H}\,|\,\phi> - <\chi\,|\,\hat{H}\,|\,\chi>)\,. \quad (11)$$

An estimate of the matrix elements is straightforward:

$$< \phi \,|\, \hat{H} \,|\, \phi > = E_{(a,b)m} - \beta\sigma H + < \phi \,|\, \Omega_B\, p_x\, y + (1/8)\, M\, \Omega_B^2\, y^2 \,|\, \phi > +$$

$$< \phi \,|\, \Omega_B\, p_y\, x + (1/8)\, M\, \Omega_B^2\, x^2 \,|\, \phi >$$

$$= E_{(a,b)m} - \beta\sigma H + (e/2c) < \phi \,|\, (p_x / M)\, y \,|\, \phi > H + ¼\, (e/2c) < \phi \,|\, \Omega_B\, y^2 \,|\, \phi > H$$

$$+ (e/2c) < \phi \,|\, (p_y / M)\, x \,|\, \phi > H + ¼\, (e/2c) < \phi \,|\, \Omega_B\, x^2 \,|\, \phi > H$$

$$= E_{(a,b)m} - \beta\sigma H + \mu_{A\phi\phi} H + \Delta\mu_{\phi\phi} H \qquad (12)$$

$$< \chi \,|\, \hat{H} \,|\, \chi > = (n_x + ½)\, \hbar\, \Omega_B + (n_y + ½)\, \hbar\, \Omega_B + p_z^2 / 2M - \beta\sigma H +$$

$$(M\omega^2 / b)\, (d_c - d_b) < \chi \,|\, \Sigma( x^4 + y^4 ) \,|\, \chi > +$$

$$(M\omega^2 / b)\, d_b r_0^4 \pm E_{JT}\, [1 \pm 2 - (E_{\alpha\beta} / E_{JT})^2 ]$$

$$= (n_x + ½)\, \hbar\, \Omega_B + (n_y + ½)\, \hbar\, \Omega_B + p_z^2 / 2M - \beta\sigma H +$$

$$(M\omega^2 / b)\, d_b r_0^4 \pm E_{JT}\, [1 \pm 2 - (E_{\alpha\beta} / E_{JT})^2] +$$

$$(M\omega^2 / b)\, (d_c - d_b)\, \{ x_0^4 + y_0^4 +$$

$$[(n_x + ½)\, \hbar / (M\Omega_B)]\, [\, [(n_x + ½)\, \hbar / (M\Omega_B)] + 4x_0^2] +$$

$$[(n_y + ½)\, \hbar / (M\Omega_B)]\, [\, [(n_y + ½)\, \hbar / (M\Omega_B)] + 4y_0^2] \} \qquad (13)$$

$$< \chi \,|\, \hat{H} \,|\, \phi > = ½ < \chi \,|\, \Omega_B\, p_x\, y \,|\, \phi > + ¼ < \chi \,|\, (e\, H / 2c)\, \Omega_B\, y^2 \,|\, \phi >$$

$$= (e / 2c) < \chi \,|\, (p_x / M)\, y \,|\, \phi > H + ¼\, (e / 2c) < \chi \,|\, y^2\, \Omega_B \,|\, \phi > H +$$

$$(e / 2c) < \chi \,|\, (p_y / M)\, x \,|\, \phi > H + ¼\, (e / 2c) < \chi \,|\, x^2\, \Omega_B \,|\, \phi > H$$

$$= \mu_{A\phi\chi} H + \Delta\mu_{\phi\chi} H \qquad (14)$$

$$< \phi \,|\, \hat{H} \,|\, \chi > = (M\omega^2 / b)\, (d_c - d_b) < \phi \,|\, \Sigma( x^4 + y^4 ) \,|\, \chi >, \qquad (15)$$

provided $< \chi \,|\, \phi > = < \phi \,|\, \chi > = 0$, $< \phi \,|\, \phi > = < \chi \,|\, \chi > = 1$. Here $\Omega_B = |e|\, B / Mc$,

$$\mu_{A\phi\phi} = (e / 2c)\, [< \phi \,|\, (p_x / M)\, y \,|\, \phi > + < \phi \,|\, (p_y / M)\, x \,|\, \phi >] \qquad (16)$$

$$\sim (1 / c)\, r_0^2\, e\, \Omega_{\phi\phi}$$

$$\mu_{A\phi\chi} = (e / 2c)\, [< \chi \,|\, (p_x / M)\, y \,|\, \phi > + < \chi \,|\, (p_y / M)\, x \,|\, \phi >] \qquad (17)$$

$\sim\ (1 / c)\, r_0^2\, e\, \Omega_{\phi\chi}$

are the principal magnetic dipoles due to the off-center orbital current (Ampere) [6],

$$\Delta\mu_{\phi\phi} = \tfrac{1}{4}\,(e / c)\ < \phi \mid (\, x^2 + y^2 )\, \Omega_B \mid \phi > \qquad (18)$$

$$\sim \tfrac{1}{4}\,(1 / c)\, r_0^2\, e\, \Omega_B$$

$$\Delta\mu_{\phi\chi} = \tfrac{1}{4}\,(e / c)\ < \chi \mid (\, x^2 + y^2 )\, \Omega_B \mid \phi > \qquad (19)$$

$$\sim \tfrac{1}{4}\,(1 / c)\, r_0^2\, e\, \Omega_{\phi\chi}$$

are the gyromagnetic corrections to the principal dipoles [11]. Note that $\mu_{A\phi\phi}$ and $\Delta\mu_{\phi\phi}$ are averaged over the quantum state $\mid \phi >$ of the off-center rotator, while $\mu_{A\phi\chi}$ and $\Delta\mu_{\phi\chi}$ are off-diagonal between the two basis states $\mid \phi >$ , $\mid \chi >$ . For this reason both the principle dipoles and their gyromagnetic corrections are dependent on whether the rotator is in ground state or in excited state, that is, in which rotational band its energy falls. It is also worth noting that each of $< \phi \mid \hat{H} \mid \phi >$, $< \chi \mid \hat{H} \mid \chi >$ and $< \chi \mid \hat{H} \mid \phi >$ generate both electrostatic and magnetostatic energies, while $< \phi \mid \hat{H} \mid \chi >$ generate electrostatic energies alone. $E_{(a,b)m}$ is the eigen energy of the hindered rotator in an allowed rotational band (see Appendix I). The estimates at (16)-(19) all agree with the earlier classic conclusions based on Ampere's theorem [6].

The quantity $J = e\, \Omega_{\phi\phi}$ in eqn. (16) is the electric current across the natural cross-sectrion of the off-center ring. The intraband rotational frequency $\Omega_{\phi\phi} \equiv\ < \phi \mid \Omega_{(a,b)m} \mid \phi >$ is technically a fraction of the allowed rotational bandwidth and is thereby characteristic of the system. As long as the off-center ring is the sole accessible orbit common to reorientation and gyration, the precession frequency $\Omega_B$ should be equal to any of the allowed rotational frequencies:

$$\Omega_B = \Omega_{\phi\phi} \qquad (20)$$

In as much as $\Omega_B = eB/cM$, equation (20) may be regarded as a resonance condition for the magnetic field B in that only magnetic fields $B = (c/e)M\Omega_{\phi\phi}$ meeting equation (20) should couple to the off-center system; in that sense the magnetic field B is said to be "*quantized* "[6].

By similar reasoning, the quantity $\Omega_{\phi\chi}$ proportional to the small $\phi-\chi$ overlap is a free-to-hindered rotation exchange frequency which may be too low if not vanishing at too low a temperature. In as much as the hindered rotation accounted for by Hamiltonian (7) does include free spinning, the $\phi\rightarrow\chi$ transition should be regarded as one from a spinning particle, undergoing a parallel off-center orbital motion, to a merely free spinning particle. In other words $\phi\rightarrow\chi$ is equivalent to the off-center to on-center transition which is barrier controlled, while its on-center to off-center reverse occurs spontaneously.

## 3. Comments

The present quantum mechanical study has proved the feasibility of considerations based on the classic Ampere theorem. The derived magnetic dipoles may by all means contribute to the coupling of nonlinear oscillators, and in particular to the formation of vibronic bipolarons and to the pairing of small vibronic polarons.

In so far as eqns. (AI.2) give the *exact* eigen values at "not too large $q$", while (AI.3) show just how these eigen values arrange to form the allowed energy bands of a hindered rotator, the rotational ground and excited state energies will be within the allowed bands (AI.3) and so will the corresponding eigen states of the magnetic moments (16) – (19). As $q$ is increased, the widths of the allowed bands will tend to narrow and turn to single levels at "too large $q$"; the off-center reorientation freezing-in and the magnetic dipoles vanishing.

Nevertheless, at small though finite q and perhaps beyond the magnetic dipoles will rise in magnitude due to the increase of the rotational frequency $\Omega_{\phi\phi}$ as $\phi$ goes from the ground state $\phi_0$ across the excited states $\phi_{(a,b)m}$ of the hindered rotator. Ultimately this behavior may have an impact on the pairing and clustering of local nonlinear oscillators, as well as on their related vibronic polarons and bipolarons in carrying the electric current in Li-doped alkali halides or in layered perovskites where the magnetic dipole may prove to be an essential factor.

## Appendix I

### Mathieu's bands and periodic functions

Following Abramowitz & Stegun's Handbook [12], we assume the off-center species to undergo a rotation-like reorientational motion, as described by Mathieu's equation

$$d^2Y/dv^2 + [a_r(q) - 2q\cos(2v)]\, Y(q,v) = 0 \qquad \text{(AI.1)}$$

the motion being only slightly hindered at $|q|$ "not too large". Here $|q| = \surd(2\varepsilon_B / \hbar\omega)$ is Mathieu's absolute parameter, $\varepsilon_B$ is the reorientation-hindering barrier, $\omega$ is the coupled vibrational frequency (renormalized). The eigen values of Mathieu's equation are $a_0(q)$ though $a_\lambda(q)$ and $b_\lambda(q)$ at $\lambda = 1,2,3,\ldots$ increasing absolutely with $\lambda$ where

$$a_0(q) = -(1/2)q^2 + (7/128)q^4 - (29/2304)q^6 + (68687/18874368)q^8 + \ldots$$

$$a_1(-q) \equiv b_1(q) = 1 - q - (1/8)q^2 + (1/64)q^3 - (1/1536)q^4 - (11/36864)q^5 +$$
$$(49/589824)q^6 - (55/9437184)q^7 - (83/35389440)q^8 + \ldots$$

$$b_2(q) = 4 - (1/12)q^2 + (5/13824)q^4 - (289/79626240)q^6 + (21391/458647142400)q^8 + \ldots$$

$$a_2(q) = 4 + (5/12)q^2 - (763/13824)q^4 + (1002401/79626240)q^6 -$$

$(1669068401/458647142400)q^8 + \ldots$

$a_3(-q) \equiv b_3(q) = 9 + (1/16)q^2 - (1/64)q^3 + (13/20480)q^4 + (5/16384)q^5 -$

$(1961/23592960)q^6 + (609/104857600)q^7 + \ldots$

$b_4(q) = 16 + (1/30)q^2 - (317/864000)q^4 + (10049/2721600000)q^6 + \ldots$

$a_4(q) = 16 + (1/30)q^2 + (433/864000)q^4 - (5701/2721600000)q^6 + \ldots$

$a_5(-q) \equiv b_5(q) = 25 + (1/48)q^2 + (11/774144)q^4 - (1/147456)q^5 + (37/891813888)q^6 + \ldots$

$b_6(q) = 36 + (1/70)q^2 + (187/43904000)q^4 - (5861633/92935987200000)q^6 + \ldots$

$a_6(\mathrm{q}) = 36 + (1/70)q^2 + (187/43904000)q^4 + (6743617/92935987200000)q^6 + \ldots$

$a_\mathrm{r} \equiv b_\mathrm{r} = \mathrm{r}^2 + [1/2(\mathrm{r}^2-1)]q^2 + [(5\mathrm{r}^2+7)/32(\mathrm{r}^2-1)3(\mathrm{r}^2-4)]q^4 +$

$[(9\mathrm{r}^4+58\mathrm{r}^2+29)/64(\mathrm{r}^2-1)5(\mathrm{r}^2-4)(\mathrm{r}^2-9)]q^6 \quad (\mathrm{r} \geq 7) \qquad \text{(AI.2)}$

These eigen values give rise to rotational energy bands as follows (see Figure 1):

$(a_0,a_1)$, $(b_1,b_2)$, $(a_2,a_3)$, $(b_3,b_4)$, ... for $q < 0$ (upper energy branch "+" in eqn.(9))

$(a_0,b_1)$, $(a_1,b_2)$, $(a_2,b_3)$, $(a_3,b_4)$, ... for $q > 0$ (lower energy branch "–" in eqn.(9)) (AI.3)

Under the same condition of "not too large $q$" the respective eigen states read:

$\mathrm{ce}_0(Z,q) = (1/\sqrt{2})\{1 - (1/2)q\cos(2Z) + q^2[(1/32)\cos(4Z) - (1/16)] - q^3[(1/1152)\cos(6Z) -$

$(11/128)\cos(2Z)] + \ldots\}$

$\mathrm{ce}_1(Z,q) = \cos(Z) - (1/8)q\cos(3Z) + q^2[(1/192)\cos(5Z) - (1/64)\cos(3Z) - (1/128)\cos(Z)]$

$- q^3[(1/9216)\cos(7Z) - (1/1152)\cos(5Z) - (1/3072)\cos(3Z) + (1/512)\cos(Z)] + \ldots$

$\mathrm{se}_1(Z,q) = \sin(Z) - (1/8)q\sin(3Z) + q^2[(1/192)\sin(5Z) + (1/64)\sin(3Z) - (1/128)\sin(Z)]$

$- q^3[(1/9216)\sin(7Z) - (1/1152)\sin(5Z) - (1/3072)\sin(3Z) + (1/512)\sin(Z)] + \ldots$

$\mathrm{ce}_2(Z,q) = \cos(2Z) - q[(1/12)\cos(4Z) - (1/4)] + q^2[(1/384)\cos(6Z) - (1/288)\cos(2Z)] + \ldots$

$\mathrm{se}_2(Z,\mathrm{q}) = \sin(2Z) - q[(1/12)\sin(4Z) - (1/4)] + q^2[(1/384)\sin(6Z) - (1/288)\sin(2Z)] + \ldots$

$\mathrm{ce}_\mathrm{r}(Z,q)_{p=0} \equiv \mathrm{se}_\mathrm{r}(Z,q)_{p=1}\ (\mathrm{r} \geq 3) = \cos(\mathrm{r}Z - p\pi/2) - q\{\cos[(\mathrm{r}+2)Z - p\pi/2]/4(\mathrm{r}+1) -$

$$\cos[(r-2)Z - p\pi/2]/4(r-1)\} + q2\{\cos[(r+4)Z - p\pi/2]/32(r+1)(r+2) +$$

$$\cos[(r-4)Z - p\pi/2]/32(r-1)(r-2) - (1/32)\cos[rZ - p\pi/2][2(r^2+1]/(r^2-1)^2]\} + \ldots \quad (AI.4)$$

Use is made of expansions (AI.2) through (AI.4) to calculating the relevant magnetic dipoles (16) through (19) in various states of the rotating particle. The eigen values (AI.2) and eigen states (AI.4) pertain to the band edges in (AI.3). Intermediate eigen energies and eigen functions within the allowed bands have been obtained by linear interpolation of band-edge states [8].

References

[1] For a comprehensive discussion see: The Dynamical Jahn-Teller Effect in Localized Systems. Yu.E. Perlin and M. Wagner, eds. (Elsevier Science Publishers B.V., Dordrecht, 1984).

[2] R.O. Pohl, Revs. Modern Phys. **42** (2) 201-236 (1970).

[3] M. Georgiev and J. Singh, Phys. Rev. B **58** (23) 15595-15602 (1998).

[4] G.A. Gehring and K.A. Gehring, Rep. Prog. Phys. **38**, 1-89 (1975).

[5] M. Georgiev, M. Mladenova, V. Krastev, and A. Andreev, Eur. Phys. J. B **29**, 273-277 (2002).

[6] A.G. Andreev, M. Georgiev, M.S. Mladenova, and V. Krastev, Internat. J. Quantum Chem. **89**, 371-376 (2002).

[7] L. Landau & E. Lifshits, Quantum Mechanics Part I. Nonrelativistic Theory (GITTL, Moscow, 1948), p.p. 531-536 (in Russian).

[8] P. Petrova, M. Ivanovich, M. Georgiev, M. Mladenova, G. Baldacchini, R.M. Montereali, U.M. Grassano, A. Scacco, in: Quantum Systems in Chemistry and Physics. Trends in Methods and Applications. Roy McWeeny, Jean Maruani, Yves G. Smeyers, and Stephen Wilson, eds. Topics in Molecular Organization and Engineering, Volume **16** (1997) (Kluwer Academic Publishers-Dordrecht-The Netherlands), paper 20, pp. 373-395.

[9] L. Brillouin and M. Parodi, Propagation des ondes dans les milleux periodiques. (Paris, Dunod & Masson et C^ie^, 1956).

[10] M. Glinchuk, M.F. Deigen and A.A. Karmazin, Fiz. Tverdogo Tela **15** (7) 2048 (1973).

[11] R.F. Wallis, Waves in Gyrotropic Media. Lecture given at the NATO Summer School in Sozopol (1994).

[12] M. Abramowitz and I.A. Stegun, Handbook of Mathematical Functions (Dover, New York, 1972).

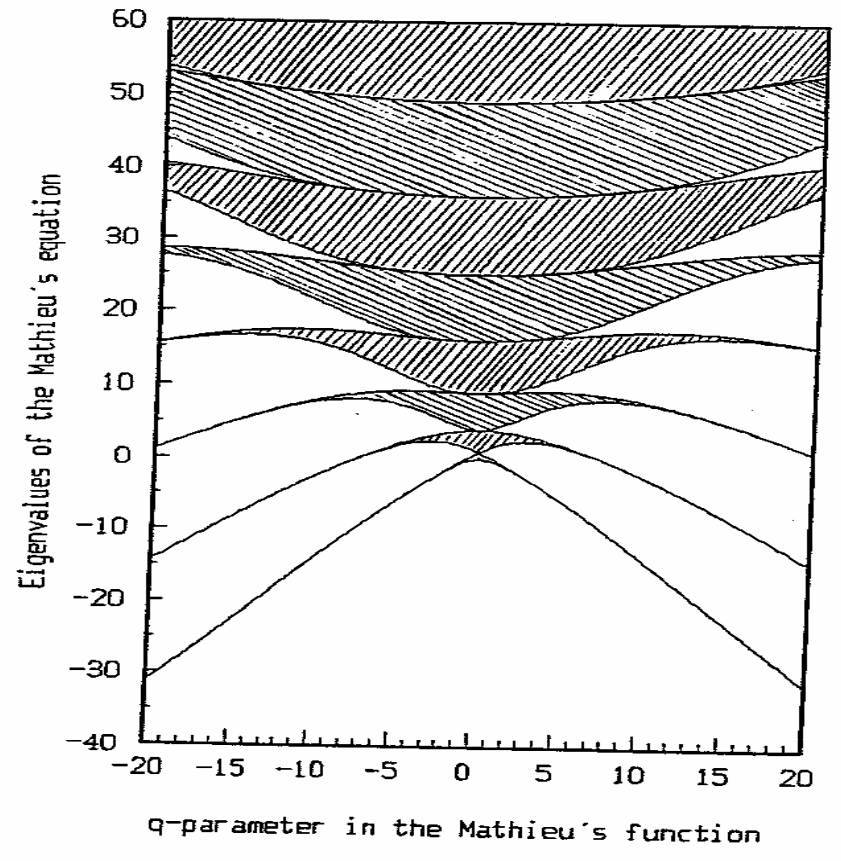

Figure 1.

Allowed rotational energy bands (hatched) of a nonlinear oscillator satisfying Mathieu's equation. The bands are seen to widen as the energy is increased and to narrow as the q parameter is raised. (Courtesy of Dr. D. Batovsky, Bangkok.) Mathieu's eigen values and eigen states are found appropriate to describing the rotational energy spectrum of a 2-D off-center ion, like the $Li^{+}$ impurity in colored KCl alkali halide.

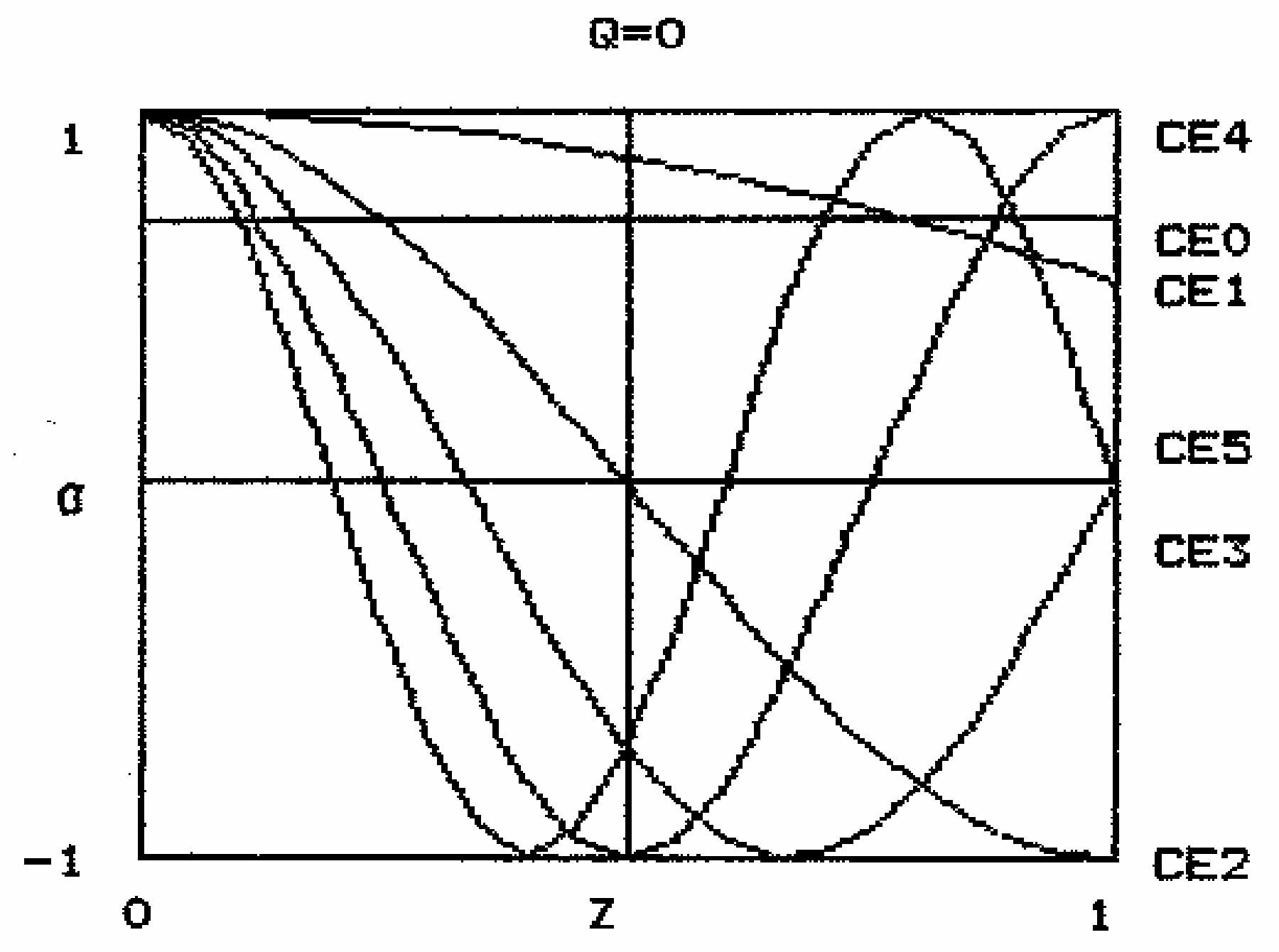

Figure 2 (a)

Figures 2 (a) through (f): The first few periodic Mathieu functions $CE_n(Z,Q)$ and $SE_n(Z,Q)$ at Q = 0, 1, 2 and n = 0 ÷ 5.

In Figures 2 (a) and 2 (b), displayed at Q = 0 are the familiar goniometric functions $CE_n(Z,0) \equiv \cos(nZ)$ and $SE_n(Z,0) \equiv \sin(nZ)$. At Q = 1 in Figures 2 (c) and 2 (d), as well as at Q = 2 in Figures 2 (e) and 2 (f), corrections are added to obtain $CE_n(Z,Q)$ and $SE_n(Z,Q)$ from the corresponding progenitor functions. See Reference [12] for details.

The above few examples will largely be utilized to compute eigen energies and other averaged nonlinear oscillator quantities.

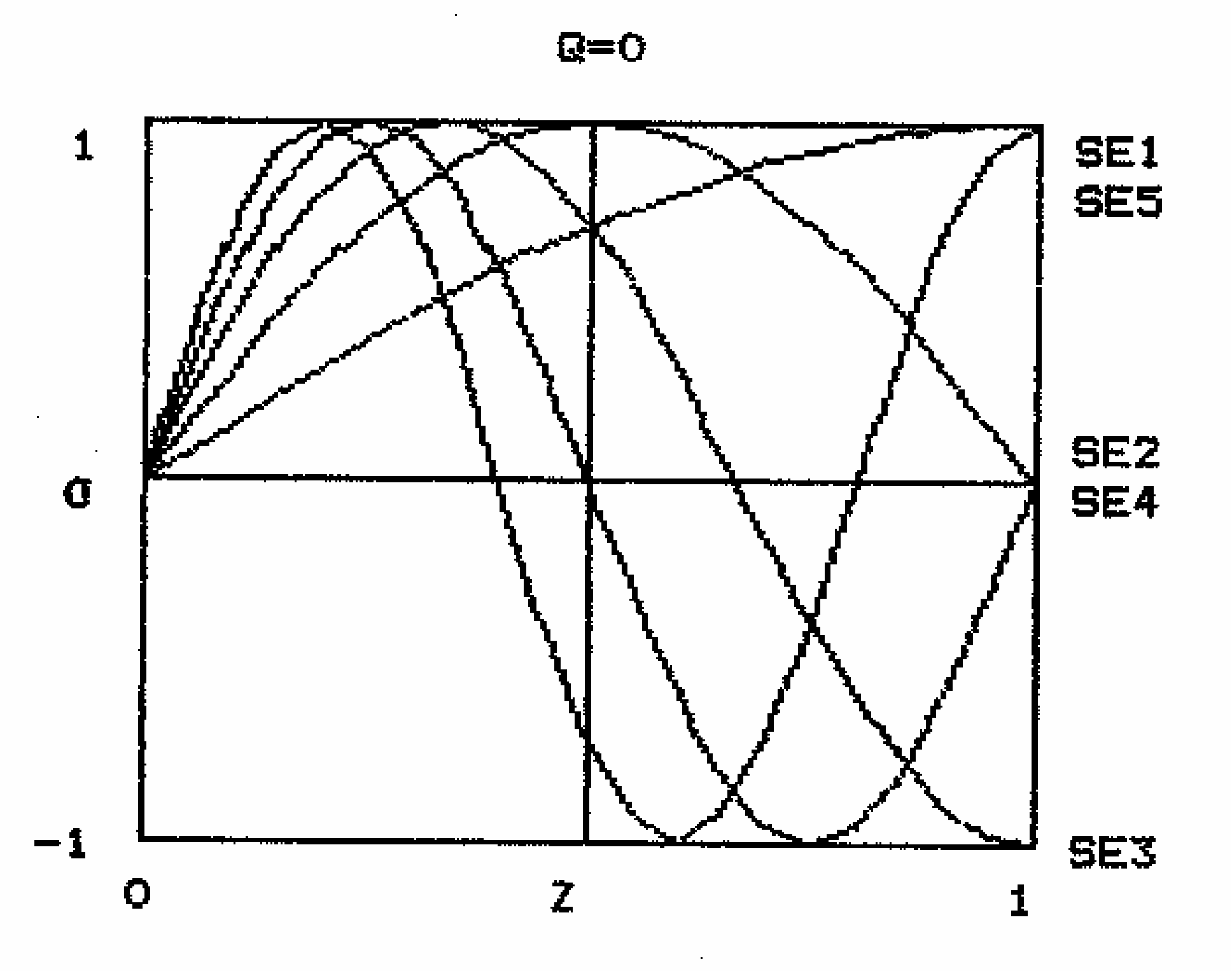

Figure 2 (b)

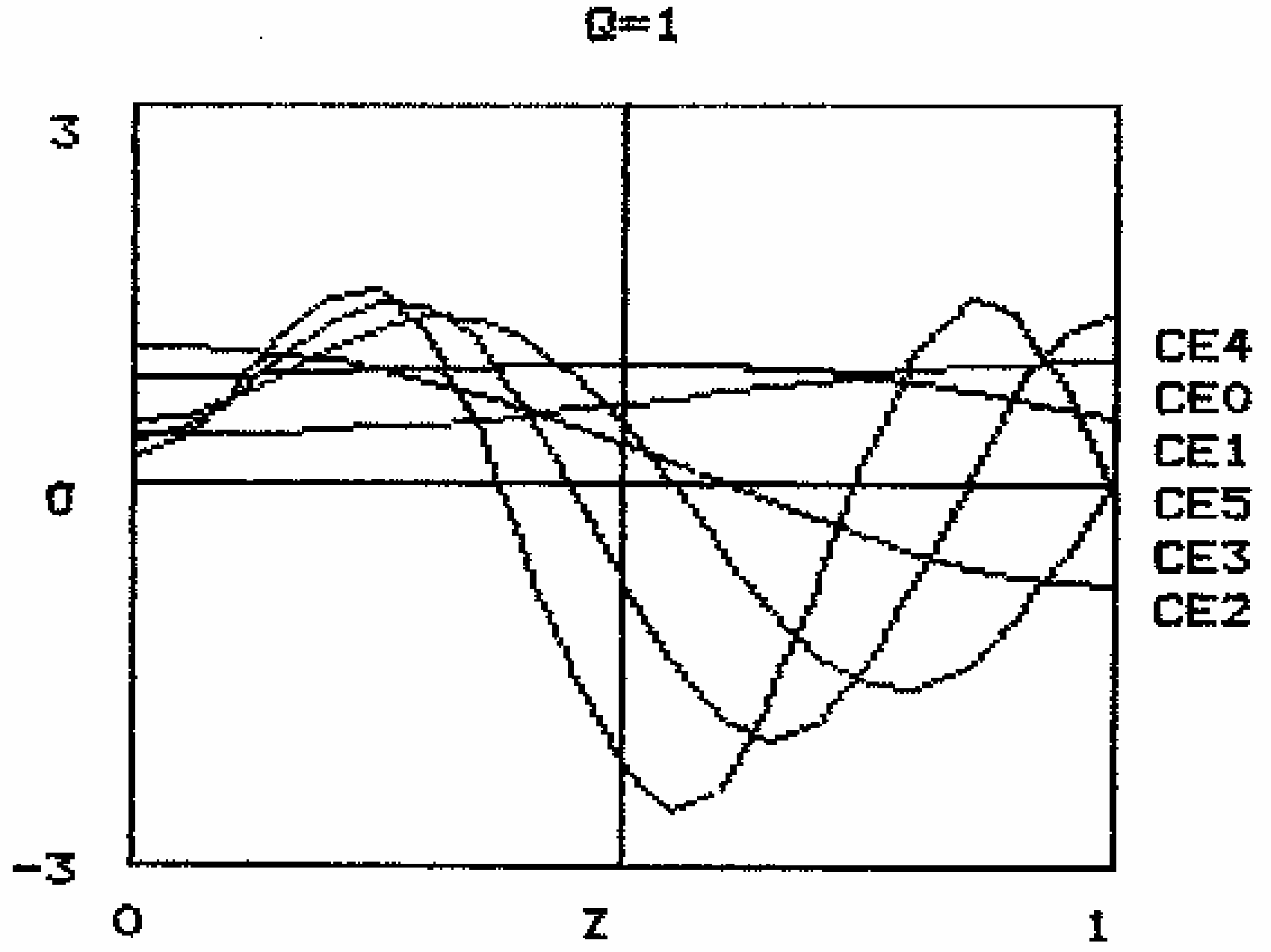

Figure 2 (c)

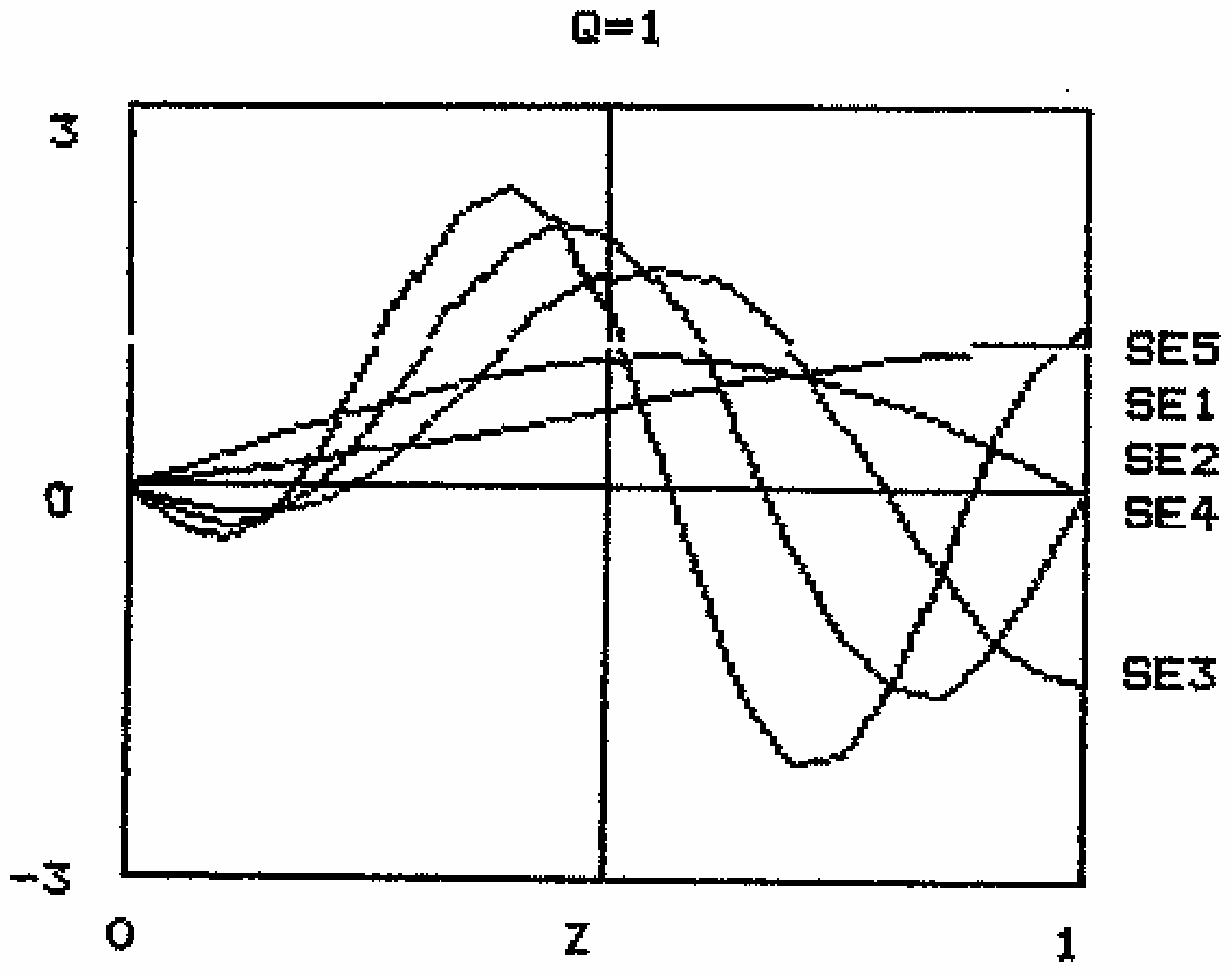

Figure 2 (d)

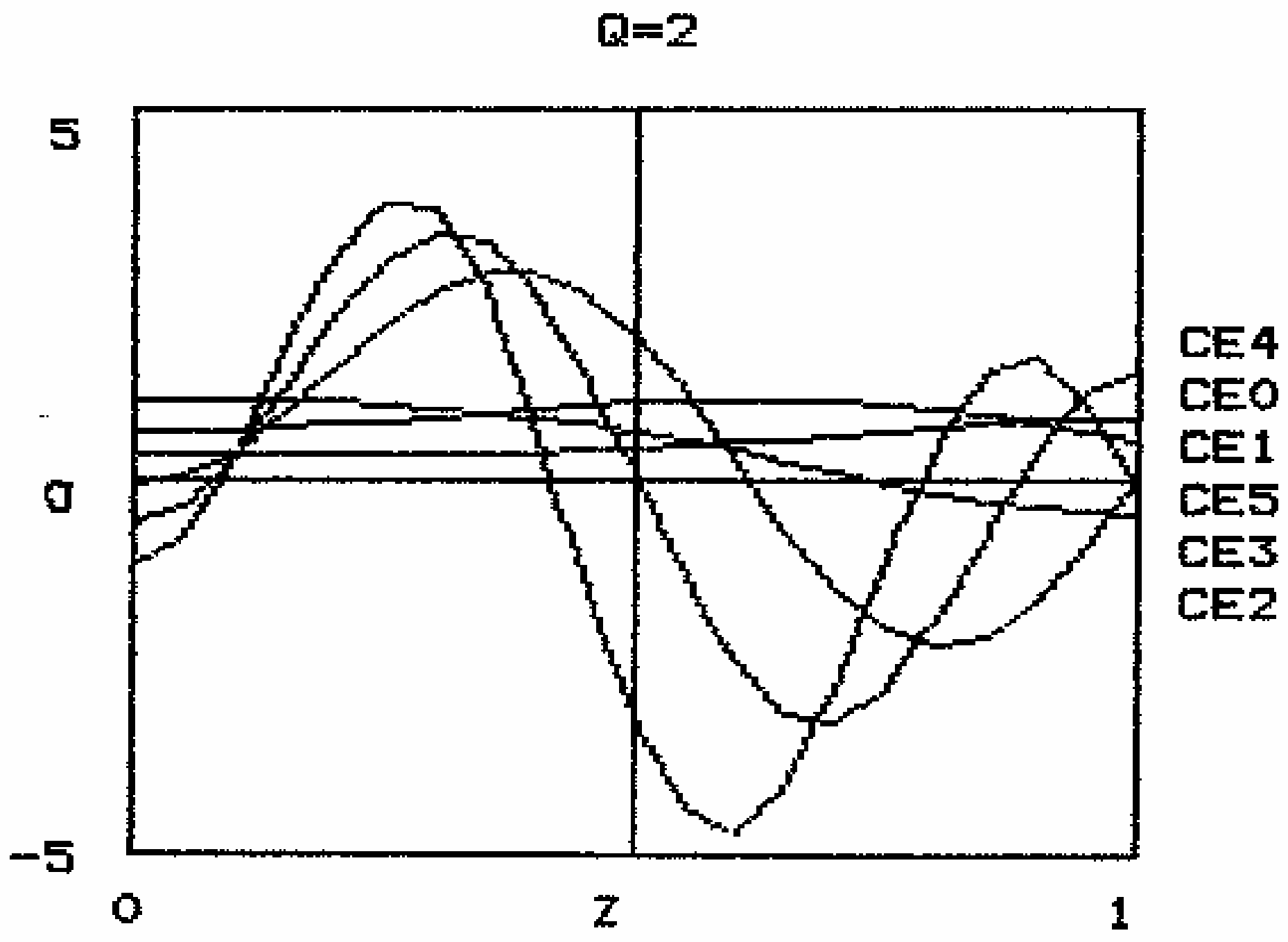

Figure 2 (e)

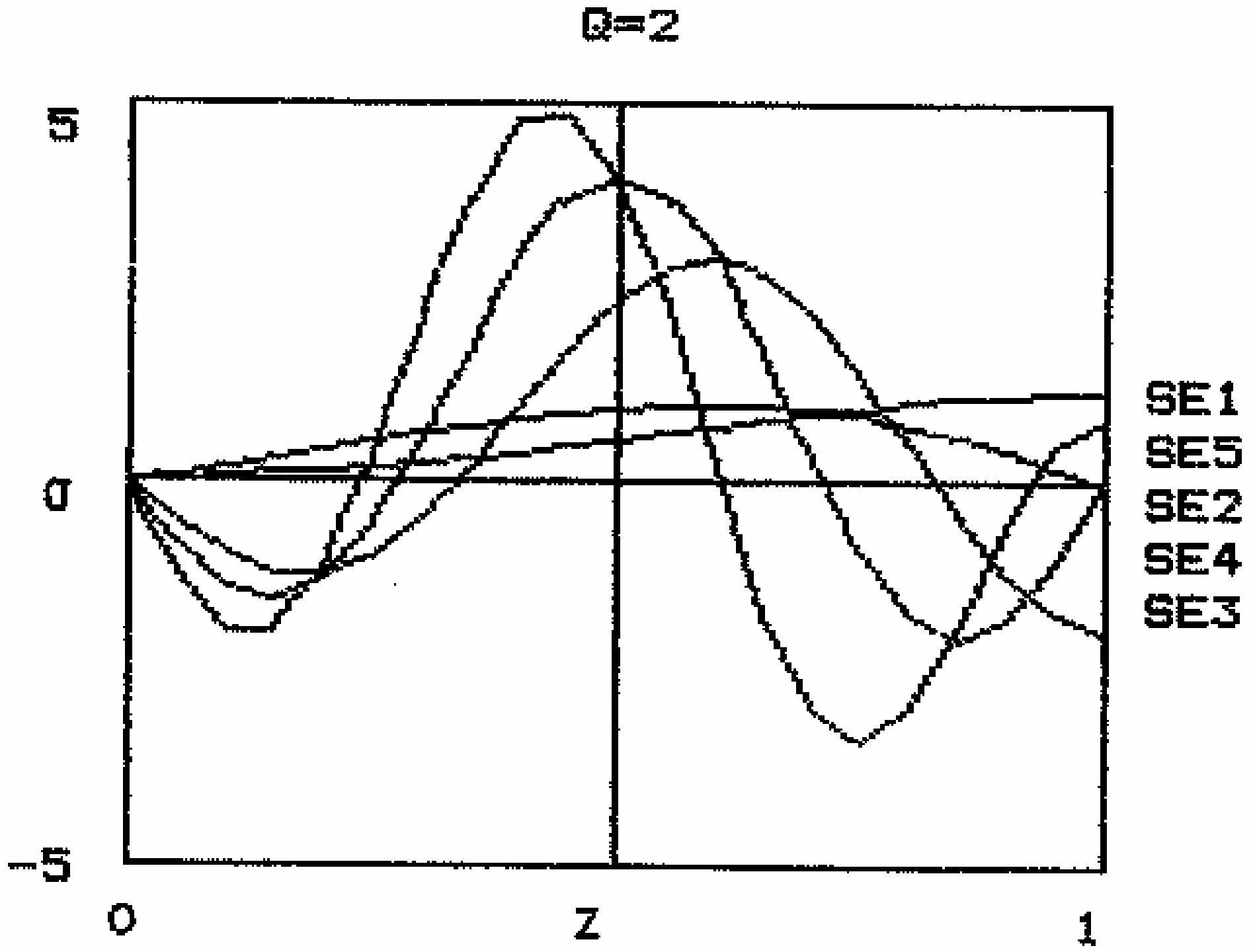

Figure 2 (f)